# Quantum Oscillations, Colossal Magnetoresistance and Magnetoelastic Interaction in Bilayered $Ca_3Ru_2O_7$

**G. Cao**

Department of Physics and Astronomy, University of Kentucky, Lexington, KY 40506

**L. Balicas, Y. Xin and J.E. Crow**

National High Magnetic Field Laboratory, Tallahassee, FL 32310

**C.S. Nelson**

Department of Physics, Brookhaven National Laboratory, Upton, NY 11973

We report magnetic and inter-plane transport properties of $Ca_3Ru_2O_7$ at high magnetic fields and low temperatures. $Ca_3Ru_2O_7$ with a bilayered orthorhombic structure is a Mott-like system with a narrow charge gap of 0.1eV. Of a host of unusual physical phenomena revealed in this study, a few are particularly intriguing: (1) a collapse of the $c$-axis lattice parameter at a metal-nonmetal transition, $T_{MI}$ (=48 K), and a rapid increase of $T_{MI}$ with low uniaxial pressure applied along the $c$-axis; (2) quantum oscillations in the gapped, nonmetallic state for 20 mK<T<6.5 K; (3) tunneling colossal magnetoresistance, which yields a precipitate drop in resistivity by as much as three orders of magnitude; (4) different in-plane anisotropies of the colossal magnetoresistance and magnetization. All results appear to indicate a highly anisotropic ground state and a critical role of coupling between lattice and magnetism. The implication of these phenomena is discussed.

PACS numbers: 72.15.Gd, 75.30. Vn, 75.50.Ee

Layered ruthenium oxides or ruthenates as a new class of correlated electron systems is characterized by the coexistence of different kinds of order and various magnetic and electronic transitions that are generally abrupt and anisotropic. These materials have increasingly drawn attention in recent years [1], but the 4d-electron based ruthenates are still by and large an uncharted territory rich with novel physical phenomena that very often deviate from those of semiconductors, metals or even 3d-electron transition metal oxides. Here the rich and intriguing physical properties observed in the bilayered $Ca_3Ru_2O_7$ provide another striking example that defies conventional notions.

$Ca_3Ru_2O_7$ is a recently discovered correlated electron system [2] that displays a low temperature Mott "insulating" transition and possesses an unusually wide array of unique physical properties [2-7]. It is characterized by an exceptionally strong coupling of charge, spin, and lattice degrees of freedom along with numerous long rang magnetic ordered phases. The intra-atomic Coulomb interaction, U, is believed to be comparable to the kinetic energy or bandwidth, W, thus, any small perturbations such as external magnetic fields can readily prompt drastic changes in the ground state. It is this competition between U and W that is at heart of the novel physical phenomena in $Ca_3Ru_2O_7$.

$Ca_3Ru_2O_7$ is known to undergo an antiferromagnetic ordering at $T_N$=56 K followed by an abrupt metal-nonmetal transition at $T_{MI}$=48 K with a partial gapping of the Fermi surface [2,4,7]. Intermediate between $T_N$ and $T_{MI}$ there exists an antiferromagnetic metallic phase (AFM) [2], which itself is a rare occurrence for a stoichiometric or undoped compound at ambient pressure [8]. $Ca_3Ru_2O_7$ also undergoes first-order metamagnetic transition to a field induced ferromagnetic metallic (FMM) phase at T<$T_{MI}$ where spins are almost fully polarized along the *a*-axis, the easy axis for magnetization below 42 K and concomitant negative magnetoresistive [2, 3]. The strong coupling between spin, charge, and lattice degrees of freedom in $Ca_3Ru_2O_7$ is manifested as well in Raman studies [4,7], where the Raman spectra reveal that the metal-nonmetal transition at $T_{MI}$ is accompanied by a softening and broadening of an out-of -phase c-axis Ru-O phonon mode and a rapid suppression of low-frequency electronic scattering associated with the formation of a partial charge gap of $\Delta_C \sim 0.1$ eV [4]. The corresponding gap ratio R=$\Delta_C/k_B T_{MI} \sim 23$ is large, suggesting that the gap, thus, $T_{MI}$ is driven by strong electronic correlations, typical of a Mott-Hubbard system [9,10]. In addition, the optical conductivity of $Ca_3Ru_2O_7$ yields a large scattering rate, leading to a remarkably short mean free path, *l*, which is highly anisotropic, ranging from 0.8 to 8 Å, well beyond the limit for bandlike transport, or the Mott-Ioffe-Regel limit, assuming a typical Fermi velocity of $10^7$-$10^8$ cm/s [5].

In this paper, we report detailed magnetic and inter-plane transport properties primarily at high magnetic fields and low temperatures. Among a wide array of interesting physical phenomena, a few major observations seen in this study are highly unusual: (1) A collapse of the *c*-axis lattice parameter unexpectedly occurs at $T_{MI}$ which increases rapidly with low uniaxial pressure applied along the *c*-axis; (2) A 90° rotation of the magnetic easy axis takes place in the vicinity of $T_{MI}$ due to a drastic change in spin-orbital coupling through the lattice degree of freedom; (3) Quantum oscillations (QOs) or the Shubnikov–de Haas (SdH) effect occurs in a state with a partial charge gap of 0.1 eV below $T_{MI}$ [4]. The QOs with extremely low frequencies are observed in the *c*-axis or inter-plane resistivity, $\rho_c$, for the magnetic field, B, parallel to the $c$-axis at low temperatures. It is also found that the QOs reoccur with a higher frequency when B is higher than the metamagnetic transition and within the $ac$-plane; (4) When B is parallel to the $a$- or $b$-axis, the inter-plane resistivity $\rho_c$ exhibits tunneling magnetoresistivity, which yields a precipitous drop in resistivity by as much as three orders of magnitude; (5) However, the decrease in $\rho_c$, or the interplane negative magnetoresistivity for B$||b$-axis, the hard-axis, is two orders of magnitude larger than that for B$||a$-axis, the easy-axis, where the magnetization is fully polarized at B>6 T [2]. All these results point toward the critical role of structural distortion and the coupling between lattice and magnetism through the orbital degree of freedom, which primarily drive the rich physical phenomena.

With a double Ru-O layered orthorhombic structure [3], $Ca_3Ru_2O_7$ belongs to the Ruddlesden-Popper series, $Ca_{n+1}Ru_nO_{3n+1}$, with n=2, where n is the number of coupled Ru-O layers in a unit cell, as shown in Fig. 1a. It has a *B2₁ma* space group (or *Cmc2₁* with the $a$-axis being the long axis) and lattice parameters of $a$=5.3720(6) Å, $b$=5.5305(6) Å, and $c$=19.572(2) Å [3]. The layered nature is clearly illustrated in the TEM image along [010] zone axis (see Fig.1b) where the stacking sequence of Ru-O and Ca-O layers is visible. The crystal structure is severely distorted by rotations and tilting of $RuO_6$ [for details see ref. 3]. The structural anisotropy in basal planes is noticeably reflected in the diffraction pattern shown in Fig.1c where extra spots due to a dynamic effect are visible along the $b$-axis. As can be seen below, this anisotropy results in a highly anisotropic ground state when B is parallel to the $a$- or $b$-axis.

Shown in Fig. 2a are the c-axis or inter-plane resistivity, $\rho_c$, and the lattice parameter for the c-axis (right scale) as a function of temperature. The data obtained from x-ray diffraction reveals a rapid decrease in the $c$-axis at $T_{MI}$, but no systematic changes in the $ab$-plane (it is not yet clear if the structural symmetry remains unchanged). The simultaneous structural, electronic and magnetic transitions at $T_{MI}$ unambiguously indicate a magnetoelastic effect, a major characteristic of $Ca_3Ru_2O_7$, which is also evidenced in the Raman study [4,7]. It seems surprising that the collapse of the $c$-axis lattice parameter, which would be expected to enhance the inter-plane orbital overlapping, does not lead to a more metallic state, but conversely, a partial gapping of the Fermi surface and nonmetallic ground state! This is evidenced by an abrupt metal-nonmetal transition at $T_{MI}$=48 K followed by a rapid increase in $\rho_c$ by as much as a factor of 18. It is likely that the collapse of the $c$-axis lattice parameter may be associated with a Jahn-Teller like distortion of the Ru-O octehedra, which lifts the degeneracy of the $t_{2g}$ orbitals by lowering the energy of the $d_{xy}$ orbital relative to that of the $d_{yz}$ and $d_{xz}$ orbitals. Recent first-principles calculations for $Sr_2RuO_4$ indicate that the shortening of the c-axis or the flattening of $RuO_6$, which results in orbital polarization and reducing bands of three $t_{2g}$ states, is the key fact to stabilize the insulating magnetic ground state [11]. This point may also be valid for $Ca_3Ru_2O_7$. As can be seen in Fig.2b, both $T_{MI}$ and magnitude of $\rho_c$ rise drastically even at low uniaxial pressure applied along the $c$-axis, for instance, at 2.5 kbar, $T_{MI}$ increases to 73 K and $\rho_c$ jumps by more than two orders of magnitude. In contrast, the uniaxial stress along the $a$-axis and hydrostatic pressure leads to a decrease in $T_{MI}$ and resistivity [12], consistent with the recent Raman results [7].

Fig.3 shows the longitudinal magnetization, M(T), as a function of temperature, T, for all three principal crystallographic axes at different magnetic fields, B. Along the $a$-axis (Fig.3a), M(T) features two phase transition temperatures, $T_N$=56 K and $T_{MI}$=48, as reported earlier [2,3]. M drops precipitously below $T_{MI}$, indicating a magnetoelastic effect occurring upon the collapse of the $c$-axis lattice parameter shown in Fig.2a. The $a$-axis is apparently the easy-axis for magnetization for T<42 K (as discussed below, the easy axis rotates to the b-axis for 42 K<T<$T_N$) since M(T) is lowest along this direction for the antiferromagnetic state. As B increases, $T_{MI}$ shifts slightly, whereas $T_N$ remains essentially unchanged initially and becomes rounded eventually. At B=6 T, the magnetic ground state becomes ferromagnetic, consistent with isothermal magnetization where the system undergoes the first-order metamagnetic transition to the ferromagnetic state with nearly fully polarized spins (saturation moment, $M_0$ >1.73 $\mu_B$/Ru), suggesting more than 85% spin polarization, assuming 2 $\mu_B$/Ru expected for an S=1 system. The metamagnetic transition along the $b$- and $c$-axis is much broader and occurs at much higher fields [2,3]. It is worth mentioning that given such a relatively low metamagnetic transition field along the $a$-axis, it is legitimate to speculate that below 42 K spins along the $a$-axis are *ferromagnetically and strongly* coupled within bilayers that are, however, *antiferromagnetically and weakly* coupled along the $c$-axis, i.e., between bilayers. The weak antiferromagnetic coupling between the neighboring bilayers is reflected by the fact that only 6T is needed to drive the metamagnetic transition and flip the spins. Shown in Fig.3b is M(T) for the $b$-axis which exhibits no anomaly corresponding to $T_{MI}$ but a much stronger peak at $T_N$. Unlike $T_N$ for the $a$-axis, $T_N$ for the $b$-axis decreases as B increases. The magnetic ground state remains antiferromagnetic. M(T) for the $c$-axis displays the two transitions which, however, are largely weakened and yet insensitive to B (Fig.3c). The different trends along the different axis suggest highly anisotropic magnetic exchange coupling that is coupled with the lattice degree of freedom.

Such a large magnetocrystalline anisotropy also leads to anomalous angular dependence of M. Shown in Fig.4 is M (B=6.2 T) as a function of angle, $\Theta$, defined as the angle between B and the easy axis, i.e., the $a$-axis. Unlike magnetization of ordinary ferromagnets such as $CrO_2$ and $SrRuO_3$ where an uniaxial magnetocrystalline anisotropy gives rise to a simple angular dependence described by $M=M_S\cos\Theta$, M for $Ca_3Ru_2O_7$ changes abruptly with $\Theta$ and shows no simple behavior similar to $M=M_S\cos\Theta$. For T<42 K, M saturates at 1.73 $\mu_B$/Ru at $\Theta=0^o$, i.e., along the $a$-axis, and precipitously drops as $\Theta$ increases and becomes 0.07 $\mu_B$/Ru at $\Theta=40^o$. As $\Theta$ further increases, there appears a broad peak around $\Theta=90^o$, i.e., the $b$-axis. When T>42 K, the angular dependence of M changes drastically as evidenced in Fig.4 $a$ and $b$. Apparently, the easy axis starts to rotate away from the $a$-axis at 42 K and finally becomes parallel to the $b$-axis at 55 K where the angular dependence follows the simple cosine-like behavior. It is noted that at 55 K where the system becomes much more itinerant, the magnetization along the $b$-axis is 0.8 $\mu_B$/Ru, only about a half of the saturation moment when the easy-axis is parallel to the $a$-axis. The reduction of the saturation moment could be due to the presence of the metallic state where electrons are expected to be much less localized than those below $T_{MI}$ but the gapping, which is driven by electron correlations, may not be reflected in the spatial behavior of the wave functions. Nevertheless, since the magnetocrystalline anisotropy is due mainly to spin-orbit coupling through lattice, the rotation of the easy axis or the change of anisotropy clearly indicates a drastic change in the spin-orbit coupling that is closely associated with the collapse of the c-axis lattice parameter or the magnetoelastic interaction. These results along with the data shown in Fig.2 suggest that the nonmetallic antiferromagnetic state below $T_{MI}$ may be indeed driven by the flattening of $RuO_6$ and the coupling of orbital and lattice degrees of freedom, as suggested in ref. 11.

Remarkably, the angular dependence of resistivity perfectly mirrors that of magnetization. Shown in Fig.5 is the angular dependence of $\rho_c$ and M for B(=6.2 T) rotating in the *ac*-plane at low temperatures. While M behaves in the same fashion as that for B rotating in the *ab*-plane, the mirrored angular dependence of $\rho_c$ (B=6.2T) confirms the unusually strong spin-charge coupling: an abrupt drop (rise) in M immediately leads to a precipitous rise (drop) in $\rho_c$ by more than an order of magnitude. The spin-charge coupling is even evidenced at $20 < \Theta < 160^o$ where a weak and broad peak of M vs. $\Theta$ results in a shallow and broad valley of $\rho_c$.

Indeed, the strong and complex coupling between spin, charge, orbit and lattice gives rise to highly anisotropic magnetic and electronic properties for different principal crystallographic directions. These are further demonstrated in the inter-plane resistivity $\rho_c$ with B parallel to the *a*-, *b*- and *c*-axis, respectively, presented below.

Fig.6a shows $\rho_c$ as a function of B for temperature ranging from 0.6 to 6.5 K as indicated. B is applied parallel to the inter-plane *c*-axis. An oscillatory component, i.e., the Shubnikov-de Haas (SdH) effect is developed as temperature decreases. The SdH oscillations at lower fields are shown in the inset for T = 20 mK in a limited field range of 2 < B < 12 T. Shown in Fig. 6b is the amplitude of the SdH oscillations as a function of inverse field $B^{-1}$ for several values of temperatures. The SdH signal is defined as ($\sigma$ - $\sigma_b)/\sigma_b$ where $\sigma$ is the conductivity (or the inverse of $\rho_c$) and $\sigma_b$ is the background conductivity. $\sigma_b$ is obtained by inverting the background resistivity, which is achieved by fitting the actual $\rho_c$ to a polynomial. The inset shows the amplitude of the SdH signal normalized by temperature in a logarithmic scale. The solid line is a fit to the Lifshitz-Kosevich formulae, $x/\sinh x$, where $x=14.69\ m_c T/B$. The fit yields a cyclotron effective mass $m_c=0.85 \pm 0.05$. Markedly, this cyclotronic effective mass is different from the enhanced thermodynamic effective mass estimated from the electronic contribution, $\gamma$, to the specific heat [2]. This apparent disagreement between the thermodynamic effective mass and the cyclotronic effective mass is quite common in heavy fermion systems [13] and is attributed to the fact that it has been difficult to resolve the higher effective masses in quantum oscillation experiments. It is clear that some portion of the Fermi surface with larger masses might have not been detected in our measurements. In addition, the SdH effect is conspicuously absent in resistivity for the $a$- and $b$-axis when B∥c.

As clearly seen in Fig.6b, the detected frequencies are extremely low, and only a few oscillations are observed in the entire field range up to 45 T (Above 25 T, no QOs are observed). Because of the limited number of oscillations, the usual fast Fourier analysis becomes unreliable in the present case. The frequencies can, however, be determined by directly measuring the period of the oscillations, for example, for $B^{-1} \leq 0.1 T^{-1}$. This analysis reveals two frequencies visible in Fig. 6b, the higher frequency of 28 T and the very low frequency of 10 T. The higher frequency of $F_1 = 28$ T, based on crystallographic data of $Ca_3Ru_2O_7$ [3] and the Onsager relation $F_0 = A(h/4\pi^2 e)$ ($e$ is the electron charge), corresponds to an area of only 0.2% of the first Brillouin zone (FBZ)! In contrast, for all compounds studied so far, the dHvA frequencies range typically from hundreds to several thousand tesla. Small orbits in $k$ space imply large orbits in real space, or in a much lower probability of circling the entire Fermi surface sheet due to any possible sources of scattering. For this reason, small orbits or low frequencies are much more difficult to be detected than higher ones, that is to say, one always detects high frequencies relatively easily, and lower frequencies with more efforts, if any, in conventional metals. The exceptionally low frequencies observed in this system are apparently intrinsic. It needs to be pointed out that the disappearance of the oscillations at B >25 T may indicate the proximity of the quantum limit (when the quantized magnetic energy $\mathcal{E} = (n + ½)he\text{B}/2\pi mc$ with n = 0 becomes comparable to the electronic or Fermi energy), which therefore imposes limitations on the applicability of the Lifshitz-Kosevich formalism [14, 15] to the data presented here. The envelope of the SdH oscillations outlined by dotted line shown in Fig.6b suggests beating between the two close frequencies ($F_2 \sim 10$ T) and a possibly warped Fermi surface. Both frequencies are likely to correspond to neck and belly orbits, respectively.

It is known that QOs are commonly observed in organic compounds and many pure metals due to the generally high purity of these materials where weak electron scattering preserves the sharpness of Landau levels. It is in general more difficult to investigate QOs in alloys and oxides where structural defects, disorder, impurities or large electron scattering inevitably broaden Landau levels and leading to a dampening of QOs. The occurrence of QOs in a partially gapped, nonmetallic oxide such as $Ca_3Ru_2O_7$ [16] would be unlikely based on conventional physics which requires the existence of a Fermi surface (a metallic state) and a long mean free path (>$10^3$ Å) for QOs to occur. It may be plausible that the apparent contradiction of observing the QOs in the nonmetallic state may be attributed to very small FS pockets resulted from the Fermi surface reconstruction at $T_{MI}$, due to the octahedral distortions discussed above and/or a spin/orbital ordering, i.e., a gap opens along most of the Fermi surface leaving small-reconstructed electron or hole pockets responsible for the QOs. It is noted that the orbital ordering is predicted to occur in the single layered $Ca_2RuO_4$ [17] where a sudden change in the basal plane at T=357 K may be associated with the orbital ordering [18,19]. For $Ca_3Ru_2O_7$, however, no similar changes in the *a*- or *b*-axis are discerned in spite of the collapse of the *c*-axis lattice parameter at $T_{MI}$ discussed above (see Fig. 2). The collapse of the c-axis lattice parameter may favor the orbital polarization [11], which, in turn, could alter the small Fermi surface facilitating the QOs. In addition, a highly inhomogeneous picture involving coexistence of large metallic and insulating clusters is widely discussed recently for manganites [20]. Within this framework, it could be speculated that SdH oscillations might occur within the metallic islands, and that the insulating clusters might cause the overall bad metallicity. However, the size of the orbits in the real space, based on the Fermi-surface estimated above, would be of the order of 1000 Å, which means the metallic clusters need to be on a near macroscopic scale. Furthermore, no QOs have ever been reported in the manganites. Should the metallic islands be responsible for the QOs, the QOs would become even stronger when the metallic state is fully recovered at high magnetic fields. As seen below, the QOs are only seen in the partially recovered metallic state but not in the fully recovered metallic state, suggesting that the QOs may be unique to the nonmetallic state. The QOs in $Ca_3Ru_2O_7$ may arguably manifest an exotic ground state, and understanding of which may require unconventional approaches and notions.

Shown in Fig.7 is $\rho_c$ as a function of B for various angles in the **ac-plane** at T=0.6 K for $0 \leq B \leq 30$ T (a) and the **bc-plane** at T=0.6 K for $0 \leq B \leq 33$ T (b). For B within the ac-plane, as B rotates away from the c-axis, the SdH effect is overpowered by a precipitous drop in $\rho_c$ or a first-order metamagnetic transition to a much more metallic state starting at $\Theta=15^o$ where $\Theta$ is defined as the angle between the c-axis and B (see Fig.7a). The critical field, $B_c$, at which the metamagnetic transition occurs, decreases as $\Theta$ decreases or as B becomes more parallel to the a-axis, along which spins are fully polarized at $B=6$ T. When $\Theta>30^o$, the SdH effect disappears completely below Bc. At $\Theta=90^o$, or B||a, the inter-plane magnetoresistivity ratio defined as $\Delta\rho_c/\rho_c(0)$ is more than 91%, much larger than the a-axis magnetoresistivity ratio for B||a , i.e., $\Delta\rho_a/\rho_a(0)= 60\%$ [21]. The larger inter-plane magnetorsesistance is believed to be due to a tunneling effect facilitated by a field-induced coherent motion of spin-polarized electrons between Ru-O planes. Because of the layered nature, the spin-polarized Ru-O planes sandwiched between insulating (I) rocksalt Ca-O planes form a array of FM/I/FM junctions engineered by nature that largely enhance the probability of tunneling and thus electronic conductivity, which depends on the angle between the moments of adjacent ferromagnets [22]. Remarkably, when B>Bc and for $20^o<\Theta<60^o$, the oscillations re-occur, showing a larger frequency of 47 T (see Fig.8 for clarity), suggesting a possible re-structured Fermi-surface after the first-order metamagnetic transition. The reoccurrence of the QOs with the larger frequency is consistent with the view that the QOs are likely due to the small Fermi-surface pockets discussed above. However, this behavior is not seen in a more metallic state when B is within the bc-plane as shown below. It is noted that the heavy mass part of the fermi-surface, which contributes to the large $\gamma$, does not change at the metamagnetic transition, i.e, $\gamma$ is the same above and below the metamagnetic transition [23]. Shown in Fig. 7b is $\rho_c$ (in logarithmic scale) as a function of B for various angles in the **bc-plane** at T=0.6 K for $0 \leq B \leq 30$ T. $\rho_c$ for B in the *bc*-plane behaves similarly to $\rho_c$ for B in the *ac*-plane, but distinct features are apparent, among them, three are remarkable: (1) The metamagnetic transitions for B in the *ac*-plane are first-order transitions with a large hysteresis whereas the transitions for B in the *bc*-plane are not only broader but also much higher than those for B in the *ac*-plane seen in Fig.7a; (2) for B in the *ac*-plane, oscillations reoccur when B>Bc, yielding a frequency of 47 T as discussed above and clearly shown in Fig.8. In contrast, no oscillations are discerned for B in the *bc*-plane even though $\rho_c$ in this configuration is smaller by as much as two orders of magnitude than that for B in *ac*-plane. The difference may suggest a critical role of the first-order metamagnetic transitions for B in the *ac*-plane, which could largely alter the fermi-surface in favor of the reoccurrence of the QOs; (3) the drop in $\rho_c$ for B||*a*, the easy-axis, is two orders of magnitude *smaller* than that for B||*b*, the hard axis, completely contrary to the anisotropy of magnetization [3,21]. At lowest temperatures, $\rho_c(0)/\rho_c(30T) \sim 10$ and $10^3$ for **B**||*a* and **B**||*b*, respectively. If the reduction of $\rho_c$ seen here is chiefly driven by the spin polarization, then it is intriguing that $\rho_c$ becomes so drastically smaller when B is parallel to the hard-axis *b* below 42 K. The spin-orbit interaction, which lowers the symmetry of atomic wave functions and mixes states, could lead to anisotropic scattering, thus, anisotropic magnetoresistance. But anisotropic magnetoresistive effects are known to be small in general. Nevertheless, the different anisotropies for magnetization and resistivity clearly points out that *spin-polarization* *alone cannot account for the colossal magnetoresistance and its large anisotropy observed here*.

A B-T phase diagram shown in Fig.9 summarizes major characteristics of $Ca_3Ru_2O_7$ when B is parallel to the *a*-, *b*- and *c*-axis, respectively. The large anisotropy as a main feature is clearly shown. When B||*a*-axis, the easy-axis, below 42 K, the colossal magnetoresistance, or the ferromagnetic metallic state (I) (FM-M (I)) is induced via the first-order metamagnetic transition at B=6 T, which is nearly temperature independent below 42 K. When B||*b*-axis or the hard-axis below 42 K, the ferromagnetic metallic state (II) (FM-M (II)) with two orders of magnitude larger conductivity is generated via the sharp transition that is strongly temperature-dependent. When B||*c*-axis (shaded area), the QOs occur in the gapped, nonmetallic state, persisting up to 6 K.

All results presented here repeatedly suggest an unusually strong coupling between spin, charge and orbit through the lattice degree of freedom that governs the ground state. This characteristic is also reflected in recent theoretical studies. As mentioned above, the theoretical analysis based on the multi-orbital Hubbard model coupled to lattice distortions predicts orbital ordering and large magnetoresistance in the ruthenates [17]. A phase diagram generated from this theoretical analysis suggests a potential competition between metallic FM and insulating AFM states. Since the two-phase competition is a key concept to understand colossal magnetorsesistance in manganites [22], the colossal magnetoresistance is also predicted to occur in the ruthenates [17]. In addition, the first-principles calculations indicate that the flattening of $RuO_6$ octehedra, or more generally, structural distortions are critical in determining the ground state in $Sr_2RuO_4$ [11]. This point seems to be also applicable in $Ca_3Ru_2O_7$ where the critical role of the c-axis shortening for the presence of the AFM nonmetallic ground state is so evident. This is consistent with the results of the recent Raman study that also illustrate the important role of $RuO_6$ octehedral flattening in stabilizing the AFM state in $Ca_3Ru_2O_7$ [7]. It is remarkable that the different kinds of ordering, which are conventionally expected to exclude each other, seem to be characteristically synergistic in $Ca_3Ru_2O_7$. The extraordinary coexistence of both the nonmetallic and fermion-quasiparticle characteristics is intriguing and calls for innovative theoretical approaches to the extended, correlated electrons; the drastically different anisotropies of the colossal magnetoresistance and the magnetization signals an unusual scattering mechanism other than the spin-scattering dominated mechanism that works marvelously well in other materials. To address these profound new problems will surely broaden and deepen our understanding of fundamentals of correlated electrons in general and the ruthenates in particular.

Acknowledgements: GC would like to thank Dr. Elbio Dagotto, Dr. Daniel Agterberg and Dr. David Singh for very useful discussions. YX was supported by NHMFL under Cooperative Agreement DMR-0084173. The microscopy facilities were supported by Magnet Science and Technology Division at NHMFL.

Captions:

Fig.1. (a) Crystal structure of $Ca_3Ru_2O_7$; (b) The TEM image along [010] zone axis; (c) The diffraction pattern of the basal plane.

Fig.2. (a) Resistivity, $\rho_c$, (left scale) and the c-axis lattice parameter as a function of temperature, T. (b) $\rho_c$ vs. T at low uniaxial pressures, 0, 1,5, 2, 2,5 kbar.

Fig.3. Magnetization, M, as a function of T for (a) the a-axis, (b) the b-axis and (c) the c-axis at various magnetic fields.

Fig.4. M at B=6.2 T as a function of angle, $\Theta$, defined as the angle between B and the easy axis, i.e., the *a*-axis for (a) T=10, 42, and 46 K and (b) T=49 and 56 K.

Fig.5. The angular dependence of (a) M and (b) $\rho_c$ at B=6.2 T for B rotating in the *ac* plane at low temperatures.

Fig.6. (a) $\rho_c$ as a function of B for temperature ranging from 0.6 to 6.5 K as indicated for B ∥ *c*-axis; (b) the amplitude of the SdH oscillations as a function of inverse field $B^{-1}$ for various temperatures. Inset: the amplitude of the SdH signal normalized by T in a logarithmic scale.

Fig.7. $\rho_c$ as a function of B for various angles in (a) the **ac-plane** at T=0.6 K for $0 \leq B \leq 30$ T and (b) the **bc-plane** at T=0.6 K for $0 \leq B \leq 33$ T. Note that the angular dependence of the transition seem to track sine function.

Fig.8. Enlarged $\rho_c$ as a function of B for various angles in the **ac-plane** at T=0.6 K (upper panel); Amplitude of the SdH oscillations as a function of inverse field $B^{-1}$ for B>Bc and $\Theta=27^o$ (lower panel).

Fig.9. B-T phase diagram summarizes major characteristics of $Ca_3Ru_2O_7$ when B is parallel to the $a$-, $b$- and $c$-axis, respectively.

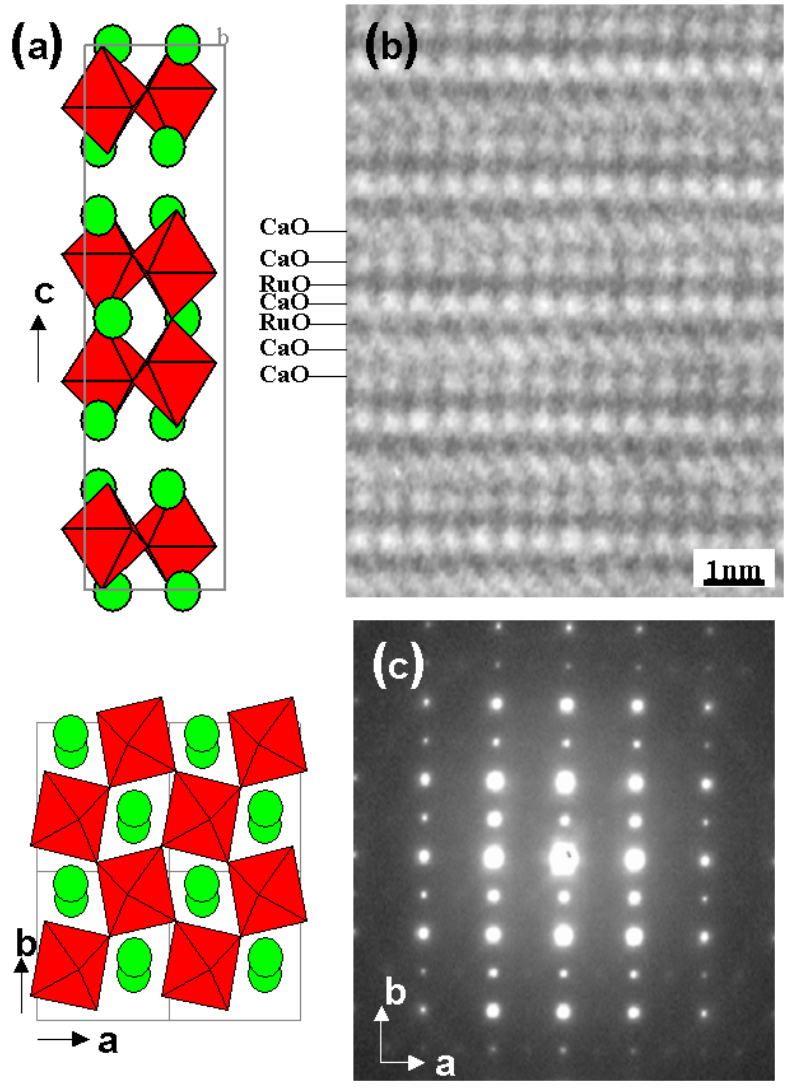

**(a)**

**(b)**

CaO
CaO
RuO
CaO
RuO
CaO
CaO

1nm

**(c)**

Fig.1, Cao, etal

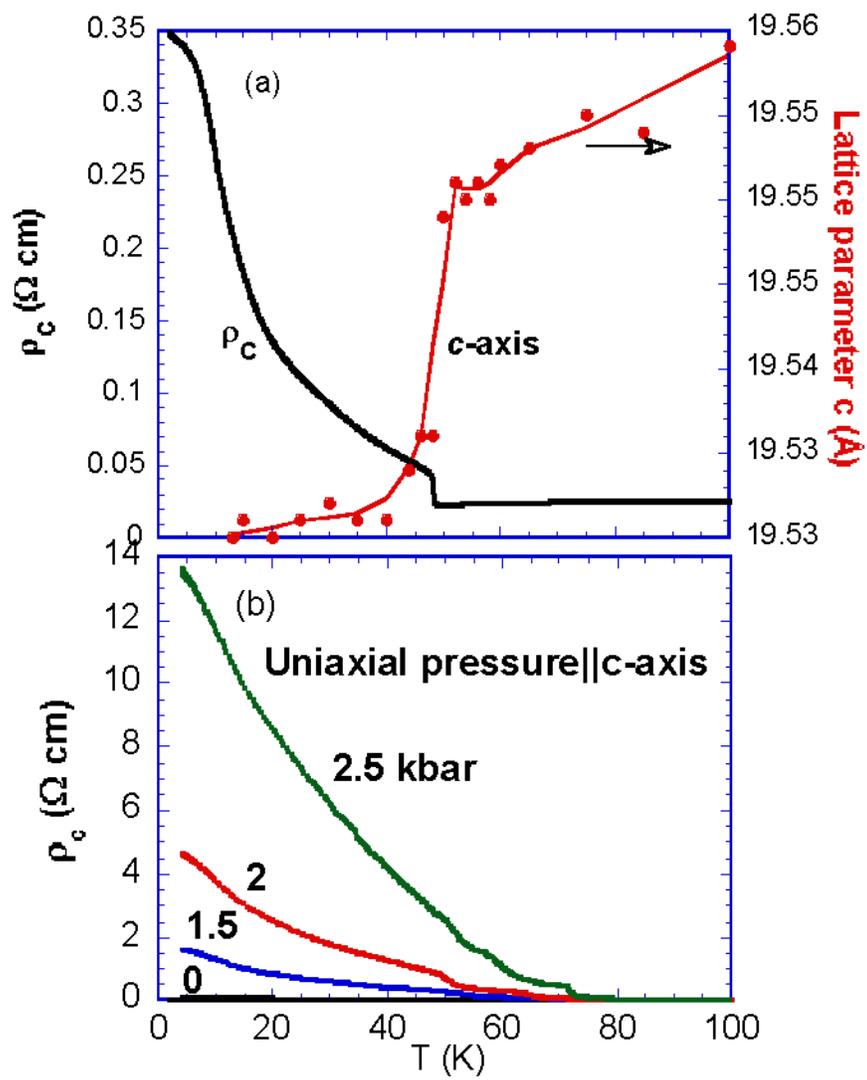

Figt.2 Cao, et al

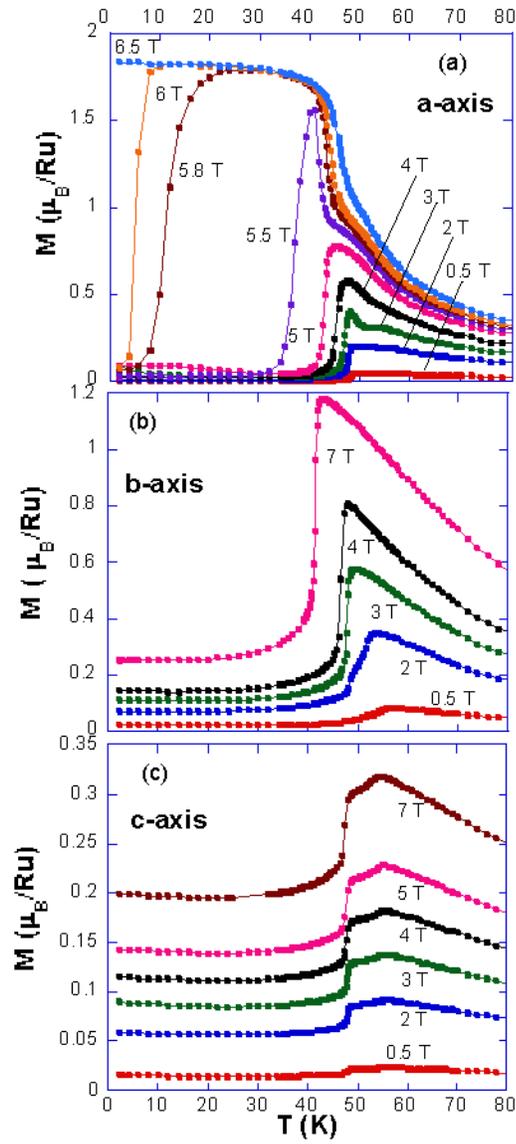

Fig.3., Cao, et al

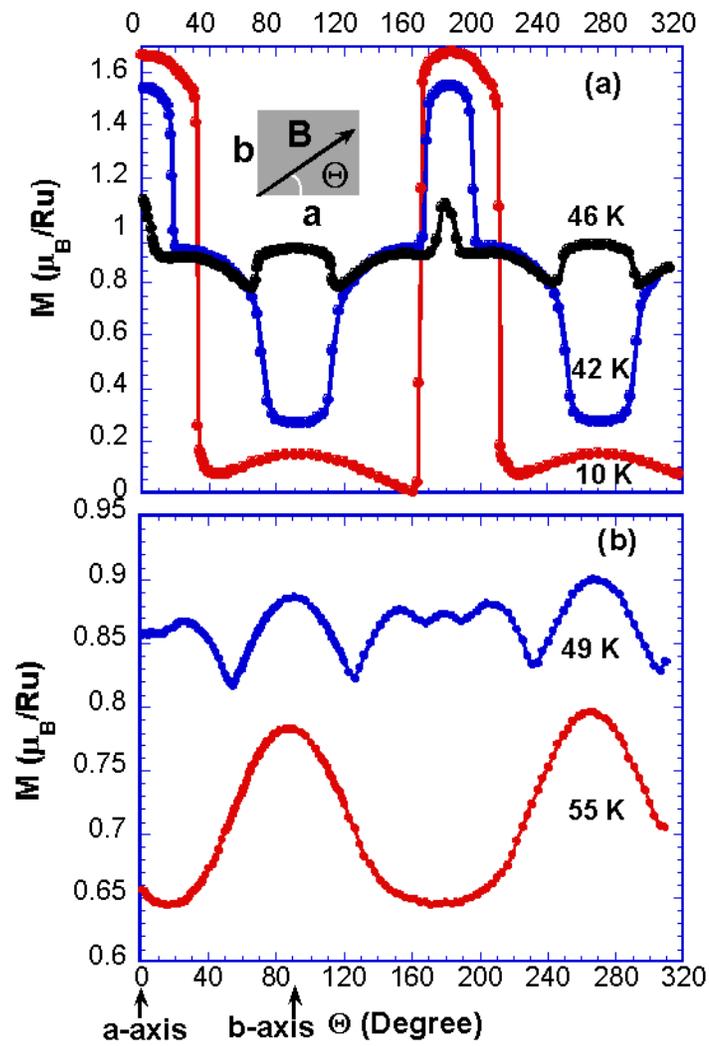

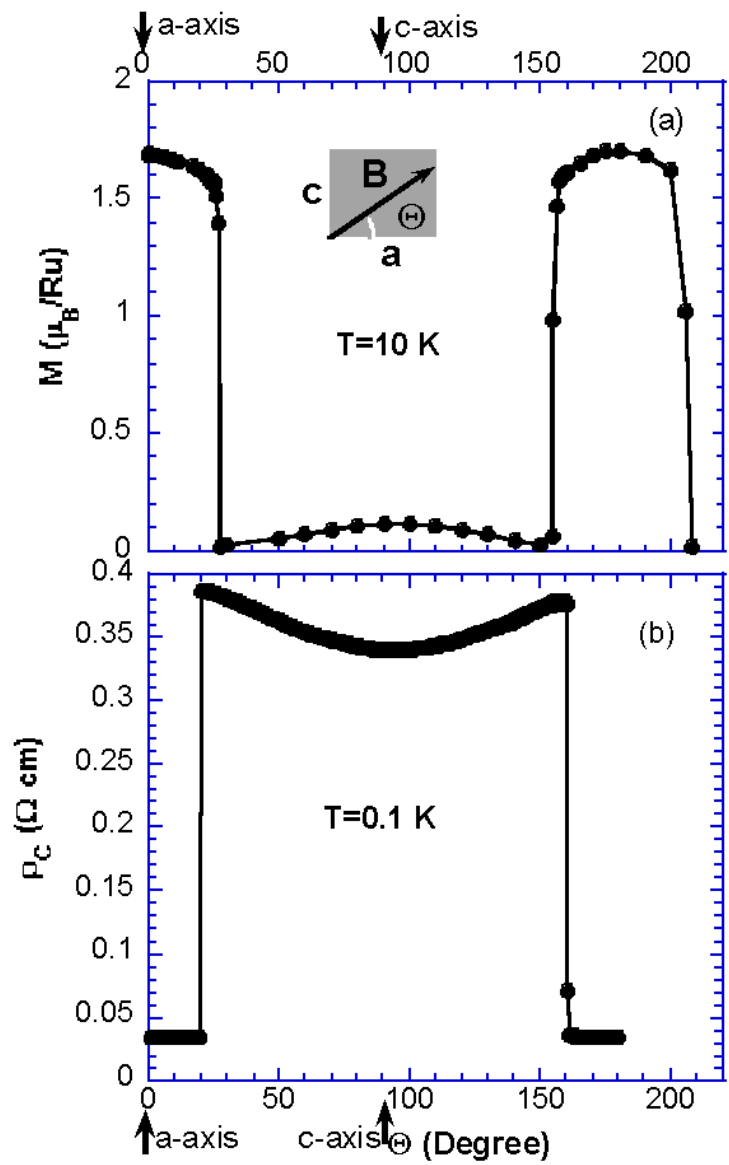

Fig.5., Cao, et al

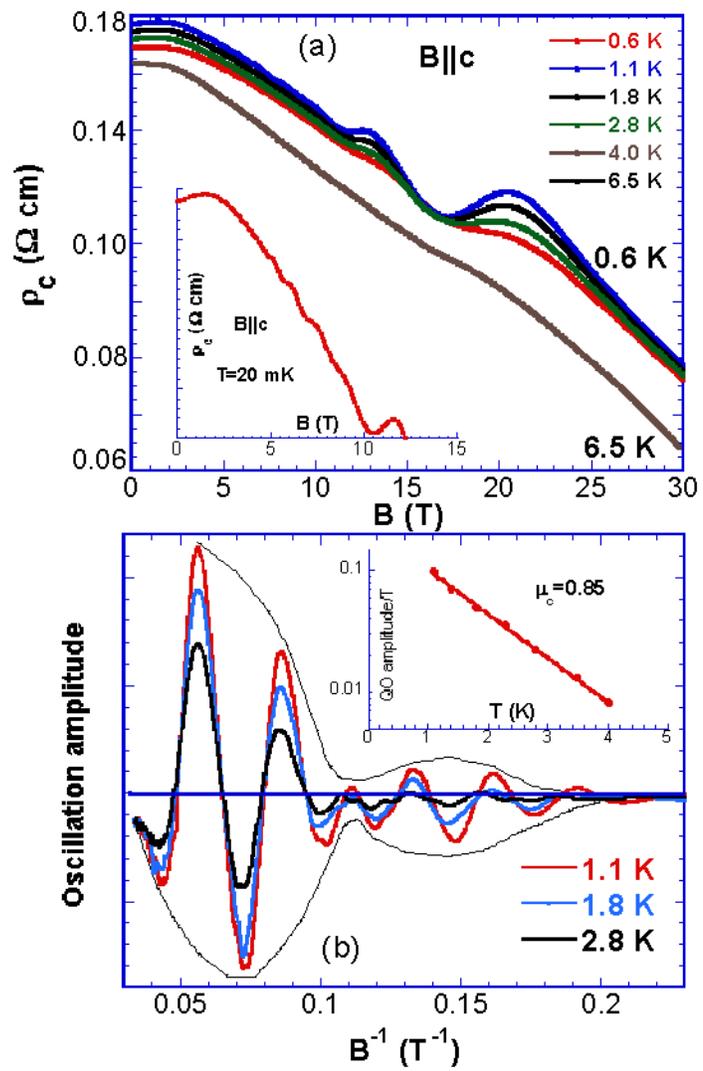

Fig.6, Cao, et al

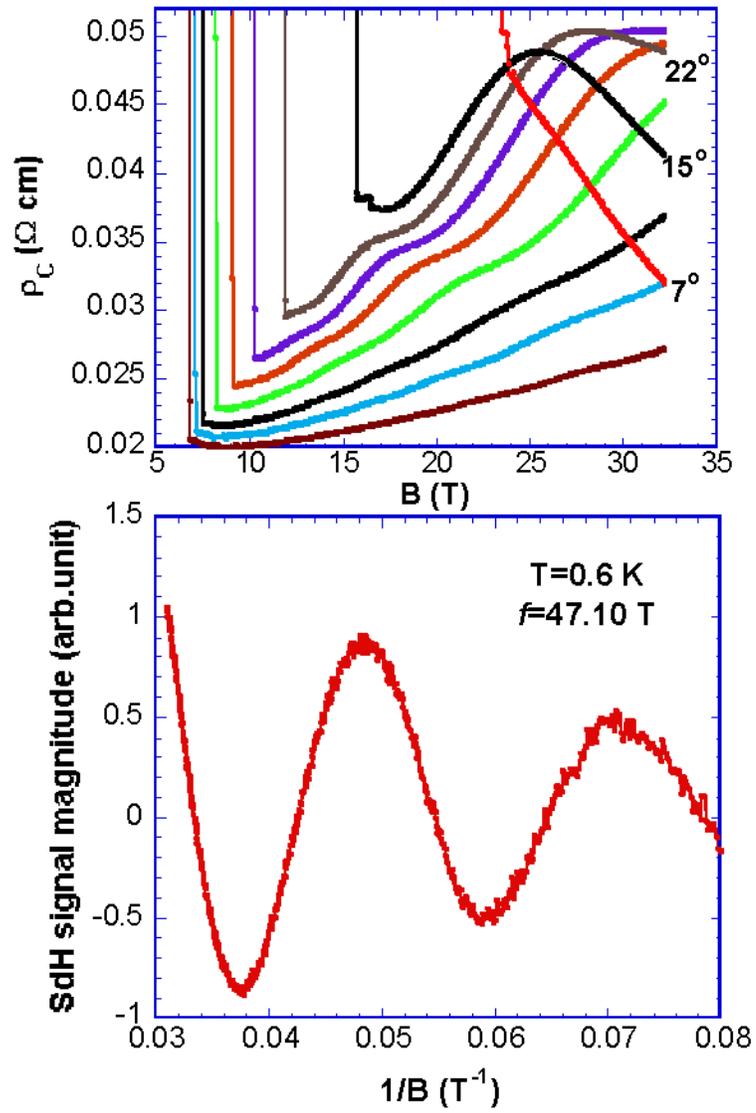

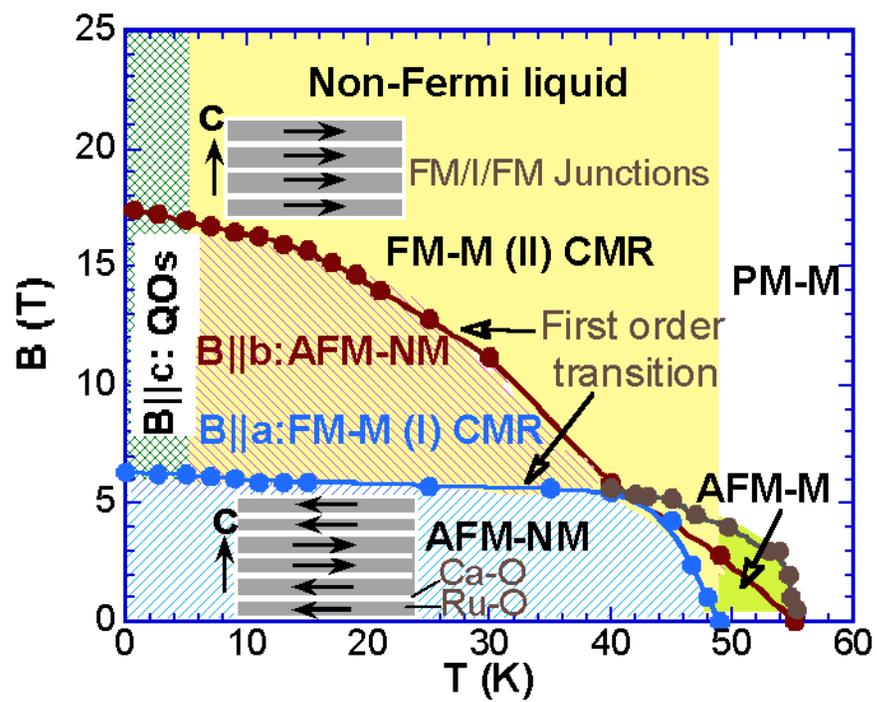

Fig.9., Cao, et al